\documentclass[aps,pre,superscriptaddress,preprint,showpacs]{revtex4-1}
\usepackage[dvipdfm]{graphicx}
\usepackage{graphicx}
\usepackage{latexsym}
\usepackage{amsmath,amssymb}
\usepackage{color}

\renewcommand{\BibitemShut}[1]{.}
\newcommand{\ie}{\textit{i.e.}}
\newcommand{\eg}{\textit{e.g.}}

\begin{document}

\title{Suppressing epidemics on networks by exploiting observer nodes}

\author{Taro Takaguchi}
\affiliation{National Institute of Informatics, 2-1-2 Hitotsubashi, Chiyoda-ku, Tokyo, 101-8430, Japan}
\affiliation{JST, ERATO, Kawarabayashi Large Graph Project, Japan}
\author{Takehisa Hasegawa}
\affiliation{Graduate School of Information Science, Tohoku University, 6-3-09, Aramaki-Aza-Aoba, Sendai, Miyagi, 980-8579, Japan}
\author{Yuichi Yoshida}
\affiliation{National Institute of Informatics, 2-1-2 Hitotsubashi, Chiyoda-ku, Tokyo, 101-8430, Japan}
\affiliation{Preferred Infrastructure, Inc., 2-40-1 Hongo, Bunkyo-ku, Tokyo, 113-0033, Japan}

\begin{abstract}
To control infection spreading on networks, we investigate the effect of observer nodes that recognize infection in a neighboring node
and make the rest of the neighbor nodes immune.
We numerically show that random placement of observer nodes works better on networks with clustering than on locally treelike networks,
implying that our model is promising for realistic social networks.
The efficiency of several heuristic schemes for observer placement is also examined for synthetic and empirical networks.
In parallel with numerical simulations of epidemic dynamics, we also show that the effect of observer placement can be assessed by the size of the largest connected component of networks remaining after removing observer nodes and links between their neighboring nodes.
\end{abstract}


\date{\today}

\maketitle

\section{Introduction}
Epidemic spreading is one of the fundamental dynamical processes that occurs on networks,
in which a node represents, for example, an individual, a computer, or an airport,
and links between them are the substrate of communication and infection~\cite{Barrat2008book,Newman2010}. 
Because it has been well known that structural properties of networks have significant impact on the consequence of epidemic spreading,
strategies for suppressing infection by the use of network structure have been widely investigated~\cite{Pastor-Satorras2001,Barrat2008book,Newman2010}. 

One of the most studied mechanisms for infection control is node vaccination,
in which a subset of nodes is chosen according to network structure and vaccinated to have perfect immunity.
Because node vaccination is equivalent to the removal of nodes from a network, the size of largest connected component after removal of vaccinated nodes is often used as a proxy measure for the effectiveness of node vaccination strategies.
Different schemes for node vaccination have been studied, including those based on node degree (\ie, the number of links connected to a node)~\cite{Albert2000,Callaway2000,Cohen2001}, node betweenness centrality~\cite{Holme2002} (see \cite{Freeman1977} for the original definition of the centrality), equal-size partitioning~\cite{Chen2008}, and community structure~\cite{Masuda2009,Salathe2010}.
Some studies considered other schemes that only use local structure, such as so-called acquaintance vaccination~\cite{Cohen2003,Holme2004}.
All node vaccination studies, both with or without knowledge of global network structure, implicitly assume that the authority distributes a finite amount of vaccine over the network and 
that the vaccinated nodes are passive in the sense that they only receive vaccination. 

Instead, one may utilize the ability of nodes to adaptively change behavior in accordance with a change in the local network in real situations such as social interactions.
Studies of another mechanism called the spread of awareness~\cite{Funk2009,Funk2010,Ruan2012,Wu2012} focus on such adaptive reactions of each node,
where a node becomes aware of infection when a neighbor node becomes infected.
When a node becomes aware of infection, the node decreases its frequency of interaction with others to reduce the chance of getting infected
and also circulates information about the infection to its neighbor nodes~\cite{Funk2009,Funk2010,Ruan2012,Wu2012}.
Similarly, in Ref.~\cite{Miller2013}, the author derived a set of deterministic ordinary differential equations which describes the time evolution of two competing epidemic processes. The author generalizes the model to the case of simultaneous diffusion of both disease and information; when a susceptible interacts with an infected or aware, the susceptible becomes aware to never get infected~\cite{Miller2013}.
A significant difference of awareness spreading from node vaccination is that there is no centralized control
and that all nodes collect the information of neighbors and adapt their own behavior.
However, it may be unrealistic to assume that all nodes have the ability to send and receive information about infection.

\begin{figure}[t]
\centering
\includegraphics[width=0.5\hsize,clip]{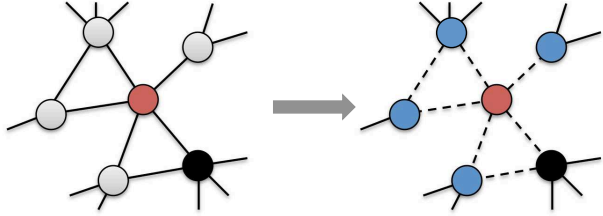}
\caption{
(Color online) Schematic of the function of an observer node. The circles filled with red (at the center), black, and blue represent an observer node, an infected neighbor, and the immunized neighboring nodes, respectively.
The dashed lines represent the links that are never used for infection, regardless of which neighbor is infected.
}
\label{fig:schematic}
\end{figure}

As a mechanism of epidemic control that takes advantage of both node vaccination (centralized and passive) and awareness spreading (decentralized and active),
we examine the placement of observer nodes on networks in this paper.
We assume that when one neighbor of an observer node becomes infected, the observer node immediately recognizes it and makes the rest of its neighbors immunize, as shown in Fig.~\ref{fig:schematic}. 
Therefore observer nodes never suffer from infection, whereas their neighboring nodes may get infected.

One may interpret this assumption in actual situations as follows. 
In online social networks, for example, users post their health condition on online communication sites so commonly that the actual number of reported cases of diseases such as flu can be predicted to some extent with the use of these posting records~\cite{Culotta2010,Paul2011,Aramaki2011,Achrekar2011}. If one assumes that an online social network is considerably overlapped with its offline counterpart, it would be a plausible intervention strategy to send alert messages to friends in the online network when one observes a post by a friend reporting their infection. In this context, placing observer nodes corresponds to persuading a fraction of users to send alert messages in case of recognizing infection in friends.
In computer networks, a computer may disconnect the computers connected to it from the rest of the network,
when the network suffers from a propagating attack such as a computer virus.
Since observation by nodes usually consumes resources,
it is important to investigate the efficacy of the protection with a given number of observer nodes
and to seek schemes for observer placement that achieve sufficient protection with a small number of observers. 

To address the two points of observer placement,
we numerically simulate an epidemic model on networks with various observer placements.
First, we investigate the relationship between network properties and the effect of observer placement when observer nodes are randomly chosen.
We then examine three heuristic schemes to choose observers based on the nodes' structural properties for synthetic and empirical networks.

\section{Results}
\subsection{Random placement of observer nodes}
We begin with a comparison of the effects of observer placement and node vaccination on a synthetic network.
As the model of spreading, we use the standard susceptible-infected-recovered (SIR) epidemic dynamics on networks~\cite{Anderson1991,Barrat2008book,Newman2010}.
In the SIR dynamics, each node takes one of three states: susceptible (S), infected (I), or recovered (R).
In the initial state, we set the states of all nodes to S, except for a single initial seed whose initial state is set to I.
Infection occurs on links between S and I nodes at rate $\lambda$ that determines the degree of infectiousness.
Infected individuals independently recover and turn to R state at rate $\mu$.
After a sufficiently long period, the population converges to a final state composed of only S and R nodes.
We set the parameter values of the SIR dynamics as $\lambda = \mu = 1$ throughout this paper unless otherwise stated.
For these parameter values, a considerable fraction of nodes get infected in the absence of observer nodes for all networks used in this paper.

\begin{figure}[t]
\centering
\includegraphics[width=0.32\hsize]{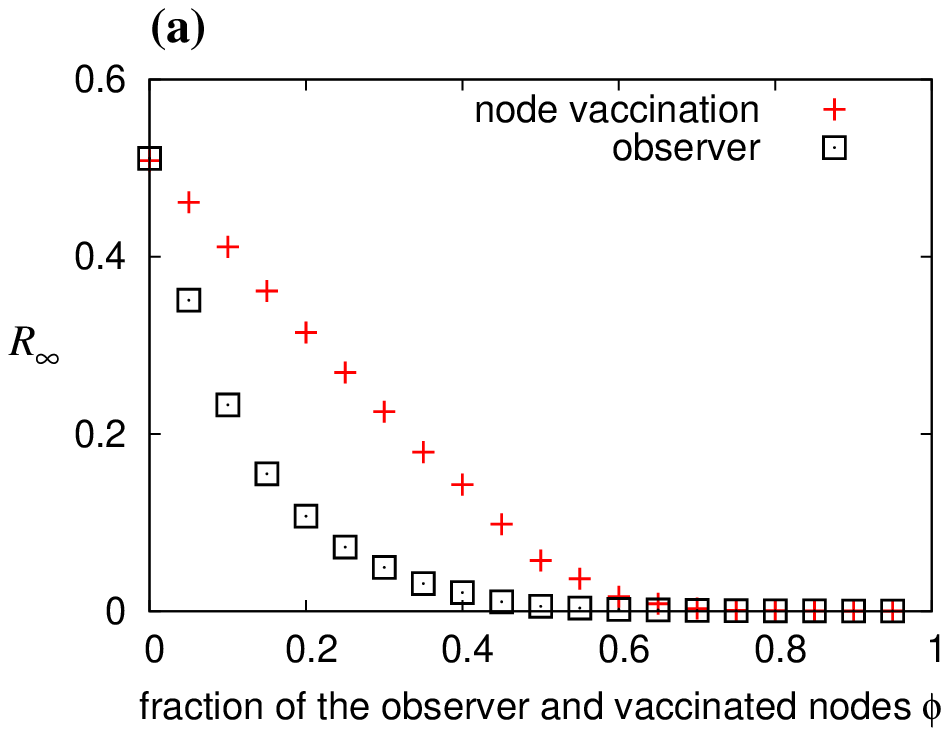}
\includegraphics[width=0.32\hsize]{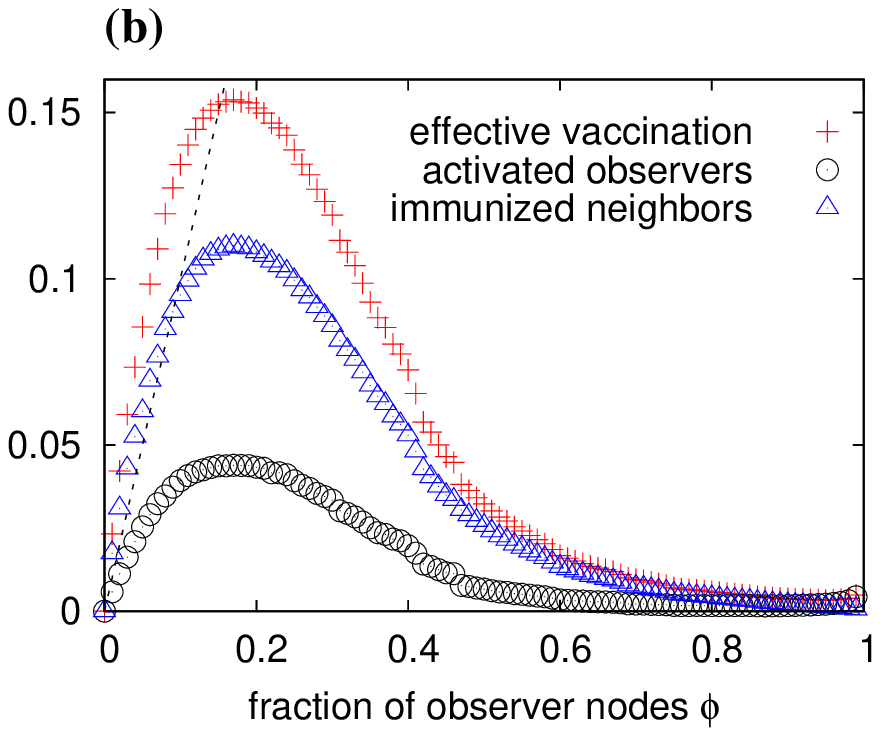}
\includegraphics[width=0.32\hsize]{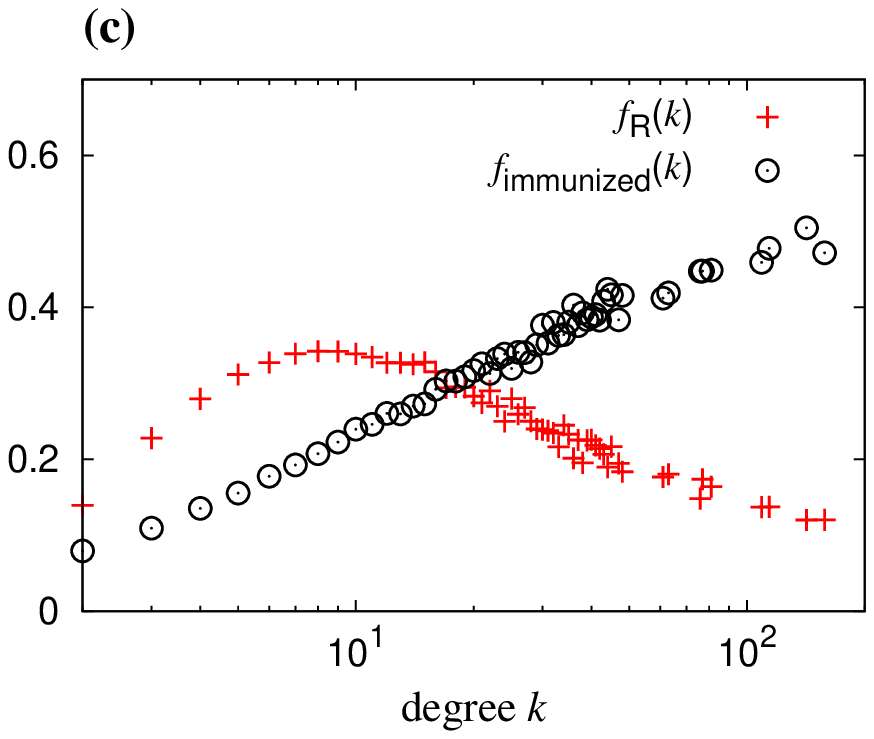}
\caption{
(Color online) Results of numerical simulations on a power-law random network with $N=10000$, $p(k) \propto k^{-3}$, and $k_{\rm min} = 2$.
(a) Final infection sizes $R_\infty$ as a function of the fraction of the treated nodes $\phi$
for random node vaccination (crosses) and random observer placement (squares),
We take the average of $R_\infty$ for $1000$ different initial seeds for each $\phi$ value over ten network instances.
(b) Fraction of effective vaccination (crosses), activated observers (circles), and immunized neighbors (triangles) as a function of the fraction of observer nodes $\phi$. The dashed line represents the diagonal line $x=y$.
We take the average of $10000$ different initial seeds for each $\phi$ value over $100$ network instances.
(c) Fraction of infected nodes $f_{\rm R}(k)$ (crosses) and nodes immunized by observers $f_{\rm immunized}(k)$ (circles) as a function of degree $k$.
We set $\phi=0.1$ and take the average of $f_{\rm R}(k)$ and $f_{\rm immunized}(k)$ for $1000$ different initial seeds on a single network instance.
}
\label{fig:R_vaccine_observer_SF}
\end{figure}

In Fig.~\ref{fig:R_vaccine_observer_SF}(a), we compare the fraction of R nodes at the final state, denoted by $R_\infty$,
for random node vaccination and the random observer placement on a power-law random network with a degree distribution $p(k) \propto k^{-3}$. We generated the power-law random network using the configuration model~\cite{Molloy1995,Newman2010}.
For node vaccination, we randomly choose a given fraction $\phi$ of the nodes to vaccinate (\ie, they behave in the same way as R nodes).
For observer placement, we randomly choose the same fraction $\phi$ of nodes.
Apparently in Fig.~\ref{fig:R_vaccine_observer_SF}(a), random observer placement considerably reduces the infection size $R_\infty$
compared to random node vaccination for the same fraction of treated nodes $\phi$.
It should be noted that $R_\infty$ for observer placement is obviously smaller than for node vaccination,
because not only are the observer nodes vaccinated but their neighboring nodes are immunized as well.
Instead, the result shown in Fig.~\ref{fig:R_vaccine_observer_SF}(a) implies 
that utilizing observer nodes can improve the effect of epidemics intervention, especially when we can control only a small fraction of nodes (\ie, small $\phi$). 

The fraction of effectively vaccinated nodes in observer placement is not so much larger than the fraction of vaccinated nodes in node vaccination.
Here, we define the fraction of effectively vaccinated nodes by the fraction of activated observer nodes and nodes immunized by activated observers (i.e., immunized neighbors).
An observer node is said to be activated if an adjacent node gets infected. 
In Fig.~\ref{fig:R_vaccine_observer_SF}(b), the fraction of effectively vaccinated nodes, activated observers, and immunized neighbors are plotted as a function of the fraction of observer nodes $\phi$.
For $0 < \phi \lesssim 0.15$, the fraction of effectively vaccinated nodes is above the line $y=x$, i.e., larger than the fraction of vaccinated nodes in node vaccination.
However, the difference is relatively small, and one cannot achieve the same effect of intervention for node vaccination
even if the same fraction of nodes are vaccinated (also see Fig.~\ref{fig:R_vaccine_observer_SF}(a)).
For $0.15 \lesssim \phi$, the fraction of effectively vaccinated nodes decreases with $\phi$,
because the initial seed finds at least one observer node in its neighboring nodes with a high probability
and the infection from the initial seed is suppressed. 

For random observer placement, 
infection is suppressed because nodes with large degrees are likely to be immunized by observer nodes.
In Fig.~\ref{fig:R_vaccine_observer_SF}(c), we plot the fraction of nodes in R state $f_{\rm R}(k)$ and nodes immunized by observers $f_{\rm immunized}(k)$
at the final state for $\phi = 0.1$ as a function of node degree $k$.
When $\phi$ is relatively small, the probability that a node has at least one observer node as its neighbor increases with $k$, 
as can be observed in Fig.~\ref{fig:R_vaccine_observer_SF}(c).
Therefore, random observer placement tends to protect nodes with large degrees and prevents infection from spreading via such hub nodes. 

We should discuss the relationship between random observer placement and acquaintance vaccination.
For acquaintance vaccination, one first choose a node randomly and then vaccinate a neighbor of the node that is also chosen randomly among all the neighbor nodes~\cite{Cohen2003,Holme2004}.
In general random networks, the probability with which a node is protected is proportional to the node's degree for both acquaintance vaccination and random observer placement and this effect prevents spreading via nodes with large degree for both strategies.
To be more precise, for acquaintance vaccination, the probability that a node with degree $k$ is protected by being chosen as a target is given by $k p(k) / \langle k \rangle$~\cite{Cohen2003}. For random observer placement, the probability that a node with degree $k$ is protected by having at least one observer in its neighbor is given by $\left( 1- (1-\phi)^k \right) \simeq k \phi$, where we assume a small $\phi$ and large $k$.
However, acquaintance vaccination would work better in random networks because the protection in observer placement is imperfect and nodes may become infected even if they have observer nodes in their neighbors.

Taking into account the results for the power-law random network (Fig.~\ref{fig:R_vaccine_observer_SF}),
one might speculate that the heavy-tailed $p(k)$ and existence of highly connected nodes are responsible for the effectiveness of this intervention,
because the protection of nodes with a large degree would be required.
Contrary to this intuition, we show that observer placement is efficient even on networks with homogeneous $p(k)$.
Another characteristic of network structure that is related to the outcome is the clustering coefficient.
The clustering coefficient $C$ of a network is defined by the average ratio of the number of triangles involving node~$i$ to the number of possible triangles $k_i(k_i -1)/2$
over all nodes, where $k_i$ represents the degree of node~$i$~\cite{Watts1998}.  
As shown in Fig.~\ref{fig:schematic}, in observer placement,
the links between the neighboring nodes of an observer as well as the links adjacent to the observer will never be used for infection.
Therefore we hypothesize that observer placement in a network with a large $C$ value is more effective than in one with small $C$ value but the same $p(k)$. 

To verify the hypothesis, we simulate the epidemic process on regular random graphs with degree $k=4$.
We vary $C$ of networks by the link-rewiring method~\cite{Maslov2002,Kim2004} as follows.
We begin with a network generated by the configuration model~\cite{Newman2010} that usually gives $C \sim 0$.
In this method, two links $(i_1, j_1)$ and $(i_2, j_2)$ are randomly chosen and rewired so as to create links $(i_1, j_2)$ and $(i_2, j_1)$.
If this rewiring increases $C$, we adopt the change. Otherwise, we discard this rewiring and choose a new pair of links.
We repeat the procedure until the $C$ value reaches a given value.

\begin{figure}[t]
\centering
\includegraphics[width=0.45\hsize]{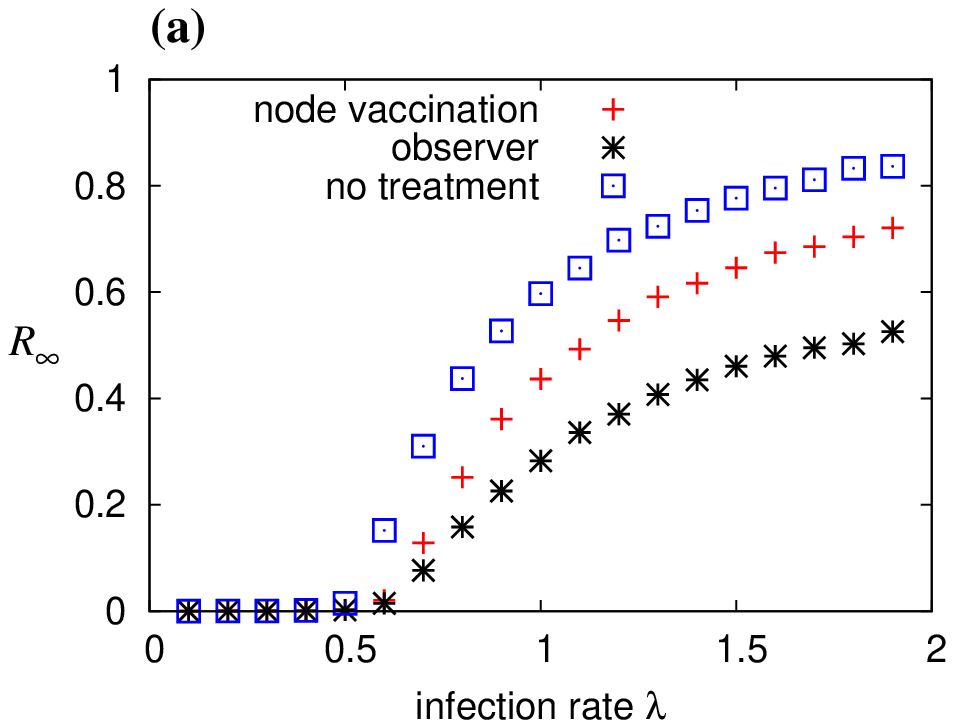}
\includegraphics[width=0.45\hsize]{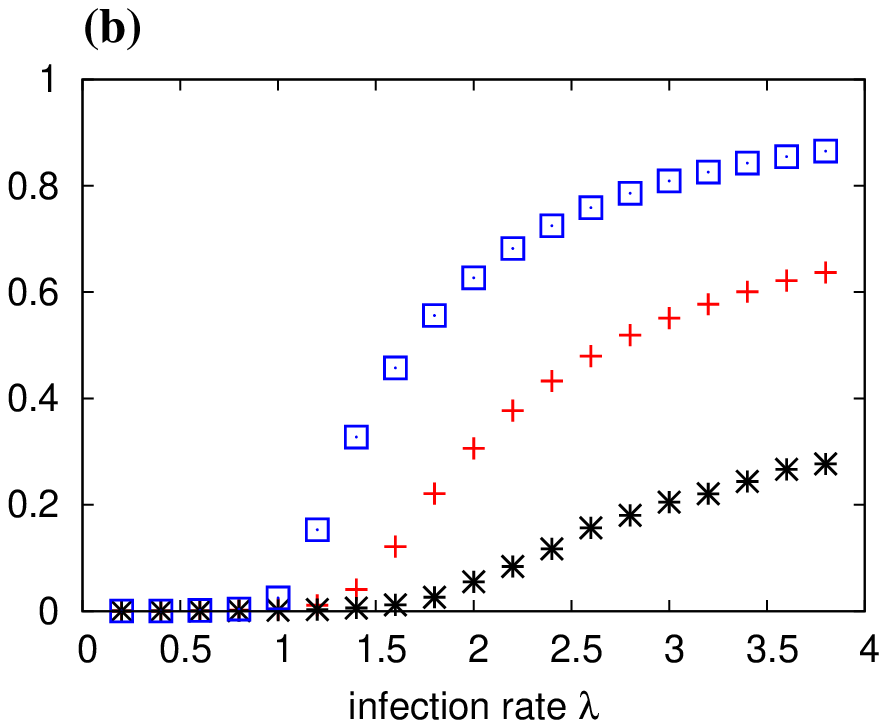}
\caption{
(Color online) Results of numerical simulations on regular random graphs with $k=4$ and $N=10000$.
Final infection sizes $R_\infty$ are shown as a function of infection rate $\lambda$
when the clustering coefficient (a) $C \sim 0$ and (b) $C=0.5$.
Three conditions are considered: random node vaccination (crosses), random observer placement (stars), and no treatment (squares).
We set $\phi = 0.1$ for both cases with $C \sim 0$ and $C = 0.5$. We take the average of $R_\infty$ for $1000$ different initial seeds over ten network instances.
}
\label{fig:R_RRG_lambda}
\end{figure}

In Fig.~\ref{fig:R_RRG_lambda} we plot $R_\infty$ as a function of infection rate $\lambda$ for $C \sim 0$ (\ie, the configuration model without link rewiring)
and $C=0.5$.
We observe that $R_\infty$ for $C=0.5$ is smaller than those with $C \sim 0$, even for no treatment or node vaccination. This is consistent with the possible effect of high clustering on the infection size as discussed in Ref.~\cite{Miller2009} (however, it should be noted that for some network models the infection size may depend on $C$ in a rather subtle manner as reported in Ref.~\cite{Newman2003}).
Nonetheless, the reduction in $R_\infty$ is the largest for observer placement.
This result implies that observer placement works well even if a network has a homogeneous degree distribution,
especially when it has large clustering. 

\subsection{Heuristic schemes for observer placement}
So far, we have considered the random placement of observer nodes and compared it with random node vaccination.
However, there should be schemes that take into account network structure and are more efficient than random placement.
Therefore we examine heuristic schemes for observer placement on the basis of node properties
and compare them on both synthetic and empirical networks.

We consider the following three schemes: degree, greedy, and ego schemes.
For the degree scheme, we choose the node with the largest degree $k_i$ to be an observer node and remove it from the network.
We repeatedly choose the node with the largest $k_i$ by recalculating $k_i$ for the remaining network at each step, until a given fraction of nodes is chosen.
For the greedy scheme, we repeatedly choose the node that has the largest number of links with unprotected nodes
(\ie, not an observer node nor the neighbor of an observer node).
It should be noted that this scheme is equivalent to the greedy approximation algorithm for obtaining the minimum dominating set of a network~\cite{Johnson1974,Raz1997}.  
For the ego scheme, we count up the number of links in the ego-centric network of each node (\ie, the network composed of the node and its neighboring nodes and the links between them).
We choose the node with the largest number of egocentric connections to be an observer node and remove the node from the network.
We repeatedly choose the nodes with the largest number of the egocentric connections in the remaining network.

\begin{figure*}
\centering
\includegraphics[width=0.32\hsize]{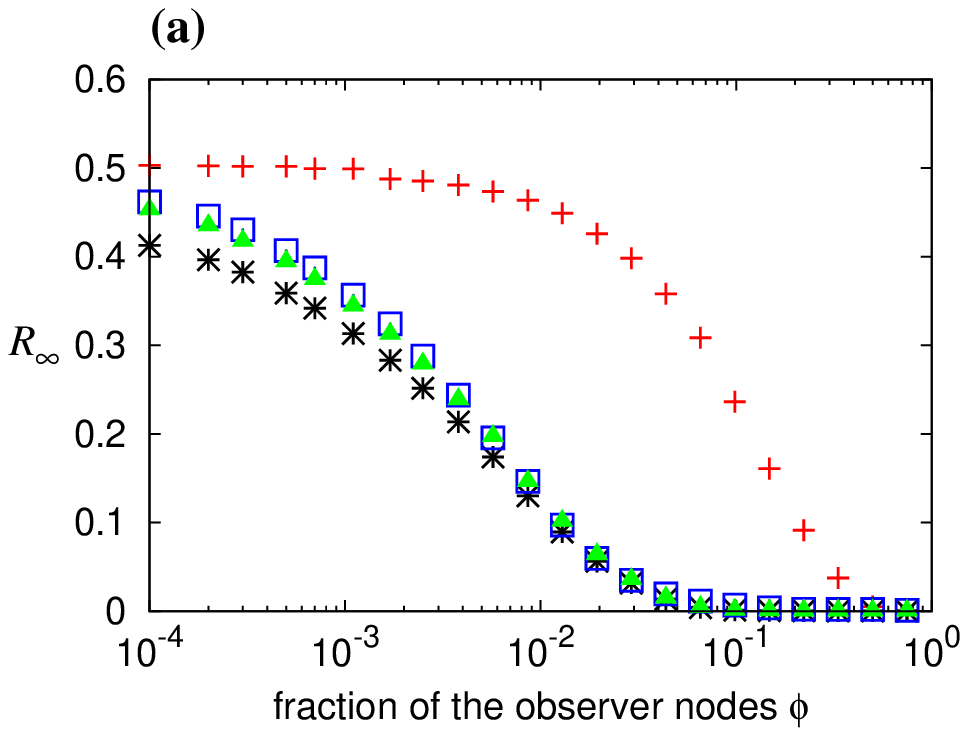}
\includegraphics[width=0.32\hsize]{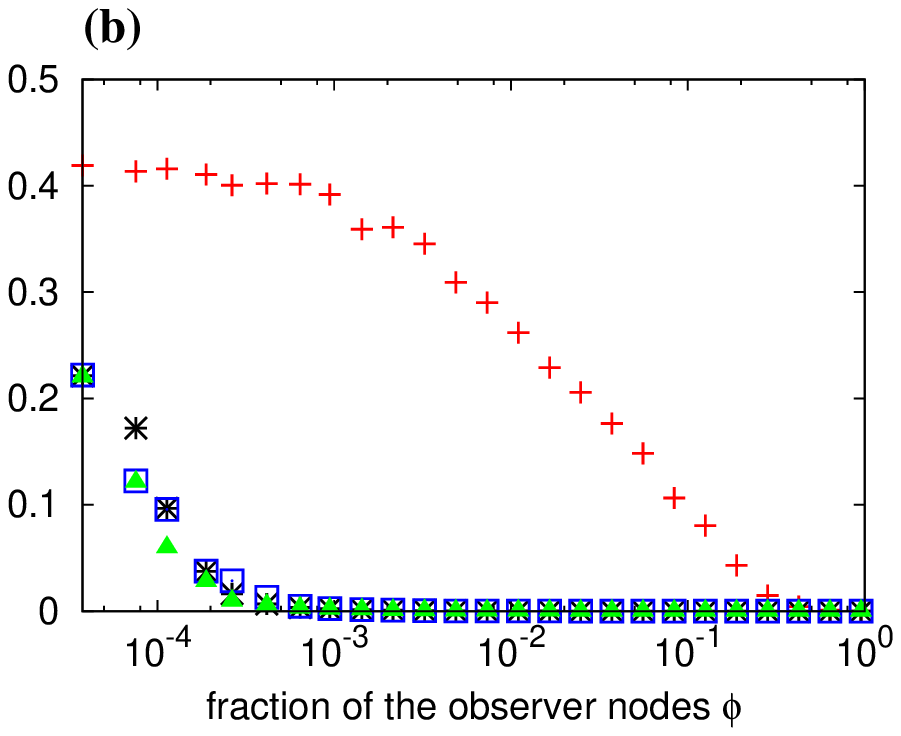}
\includegraphics[width=0.32\hsize]{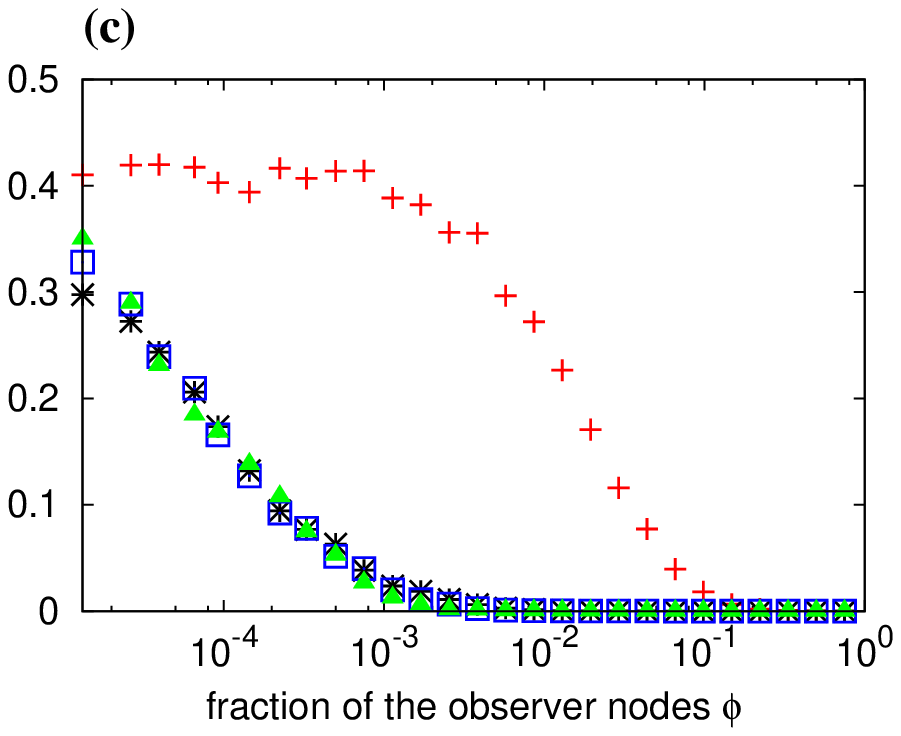}
\caption{
(Color online) Mean fraction of finally infected nodes $R_\infty$
as a function of the fraction of observer nodes $\phi$, for (a) the power-law random network with $N=10000$,  (b) the AS-CAIDA network, and (c) the Epinions network.
Four schemes for observer placement are tested: random (crosses), degree (stars), greedy (squares), and ego (triangles).
The results shown are averaged over runs with $1000$ different initial seeds for each $\phi$ for all networks. 
For the power-law random network, we average the results over ten network instances.
}
\label{fig:R_inf_heuristics}
\end{figure*}

In Fig.~\ref{fig:R_inf_heuristics} we plot $R_\infty$ for the four observer placement schemes on an SF network with $p(k) \propto k^{-3}$, the AS-CAIDA network~\cite{Leskovec2005}, and the Epinions network~\cite{Richardson2003}.
The AS-CAIDA network is an observed structure of the Internet at the autonomous system level,
and the Epinions network consists of the trust relationship between individual users of a consumer review website.
We take these two empirical networks as examples of technological and social networks, respectively.
Although the Epinions network is originally directed, we regard it as a undirected network by neglecting link direction.
As shown in Fig.~\ref{fig:R_inf_heuristics}, the random scheme is the least efficient for all three networks,
whereas the degree, greedy, and ego schemes exhibit similar performance.
The three schemes show such similar results because they choose highly overlapped sets of nodes as observers for these networks.
We confirmed that the set of observer nodes chosen in the three schemes are overlapped by $\sim 80 \%$ for the AS-CAIDA network
and $\sim 60 \%$ for the Epinions network, when $\phi$ is relatively small.
 
Although we examined the four schemes by numerically simulating epidemic dynamics,
it would be helpful if we could estimate their efficiency based on structural quantities,
especially when the network is large.
For node vaccination, previous work has employed the fraction of nodes belonging to the largest connected component (LCC)
of the remaining network after removing the vaccinated nodes, denoted by $s_{\max}$, as an indicator of the effect of vaccination schemes~\cite{Albert2000,Callaway2000,Cohen2001,Holme2002,Chen2008,Masuda2009,Salathe2010}.
In a similar way, we investigate the structural quantities that characterize the effect of observer placement.

\begin{figure*}
\centering
\includegraphics[width=0.32\hsize]{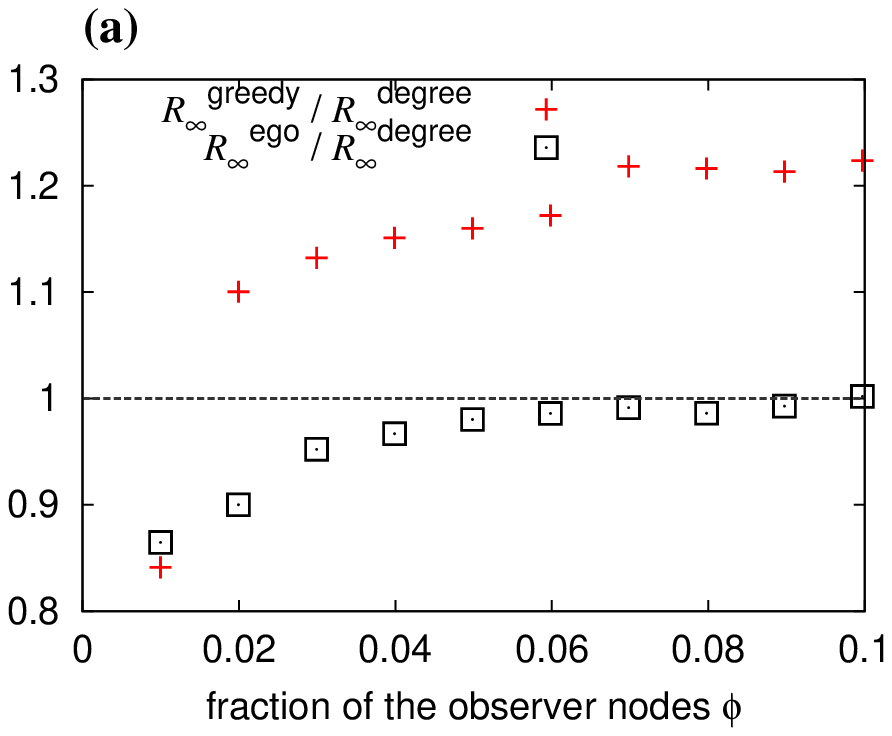}
\includegraphics[width=0.32\hsize]{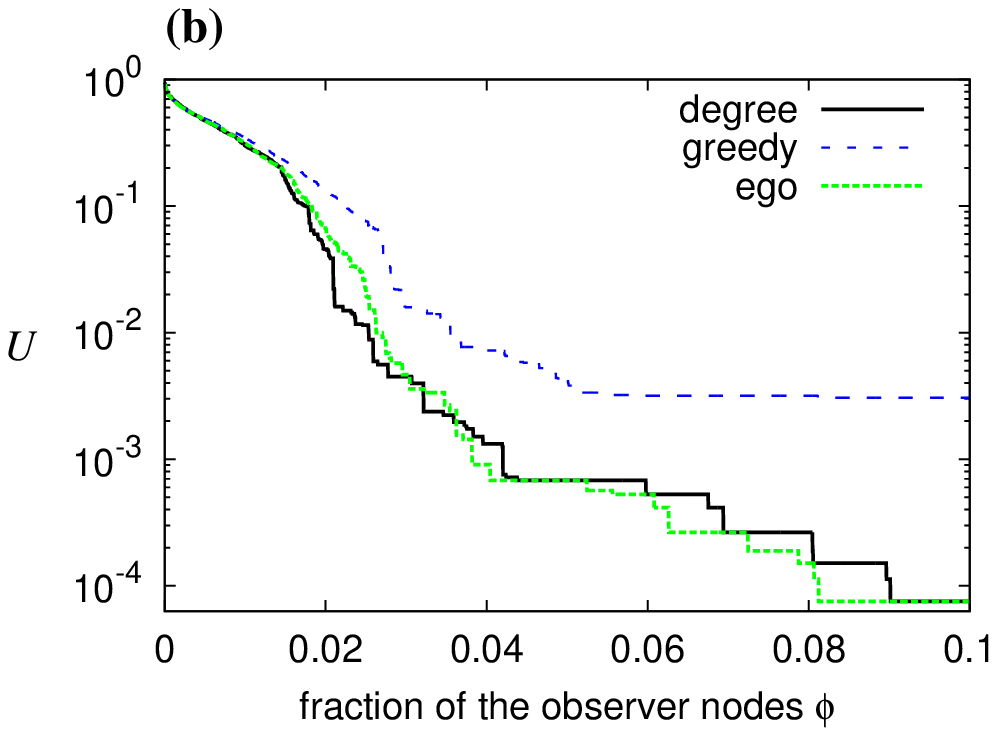}
\includegraphics[width=0.32\hsize]{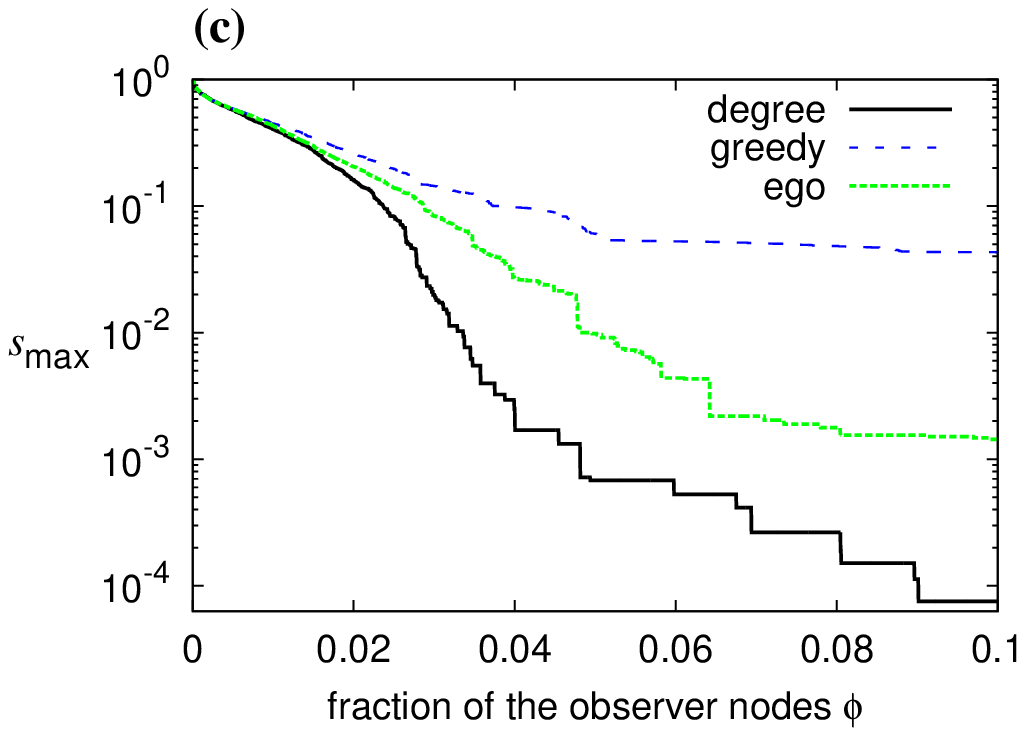}
\caption{
(Color online) (a) Ratio of the final infected nodes $R_\infty$ for the greedy (crosses) and ego (squares) schemes to that for the degree scheme
as a function of the fraction of observer nodes $\phi$ for AS-CAIDA network.
On the dashed horizontal line, the ratio is equal to unity.
(b) $U$ and (c) $s_{\max}$ as a function of $\phi$ for AS-CAIDA network with degree (solid lines), greedy (dashed), and ego (dotted) schemes.}
\label{fig:R_inf_U_S}
\end{figure*}

As a plausible measure, we focus on the fraction of nodes in the LCC of the remaining network
after removing the links in the egocentric networks of all observer nodes and denote it by $U$.
The idea of $U$ comes from the intuition that $U$ would be the largest possible size of the infection spread when placing observers.
It should be noted that $U = s_{\max}$ if the network does not have any triangles.
We show that $U$ is more useful than $s_{\max}$ to evaluate the effect of observer placement.
In Fig.~\ref{fig:R_inf_U_S}(a), we plot the ratio of $R_\infty$ for the greedy and ego schemes to the degree scheme as a function of $\phi$
for AS-CAIDA network with $\lambda =100$.
We use a very large $\lambda$ to realize the worst case of epidemic spread.
As shown in Fig.~\ref{fig:R_inf_U_S}(a), $R_\infty^{\rm greedy}$ is larger than $R_\infty^{\rm degree}$ when $0.02 \lesssim \phi < 0.1$,
whereas $R_\infty^{\rm ego}$ is only slightly smaller than $R_\infty^{\rm degree}$ when $0.05 \lesssim \phi < 0.1$.
This order in $R_{\infty}$ for the three schemes agrees with their order in $U$ (see Fig.~\ref{fig:R_inf_U_S}(b)),
while the order in $s_{\max}$ is not the same (see Fig.~\ref{fig:R_inf_U_S}(c)).
These results imply the usefulness of $U$ to estimate infection size with a particular observer placement.

In addition to the numerical results, to derive some analytical insight into the impact of network structure on $U$,
we calculate $U$ for the clustered random graph model~\cite{Newman2009,Miller2009} using generating function formalism.
As a result, we find that $U$ of random observer placement is smaller for a model with a larger $C$ value (see Appendix~A for details).
This result is consistent with our previous observation that is shown in Fig.~\ref{fig:R_RRG_lambda}.

\section{Summary}
We investigated the effect of observer nodes on suppressing epidemic spreading in networks.
We numerically showed that random observer placement works well when networks have heterogeneous degree distributions or a large clustering coefficient.
Because these two structural properties are common in social networks~\cite{Newman2010},
our results may suggest the effectiveness of observer placement in social situations.
In this paper, we mainly considered networks without well-defined communities (\ie, tightly connected subgroups).
Observer placement would take advantage of such a community structure,
because an observer node in a community will protect neighbor nodes by utilizing locally dense links
to prevent infection from escaping from (or entering) the community.
As far as we considered, the heuristic schemes for observer placement are of almost the same level of performance for both synthetic and empirical networks.
Consideration of more realistic settings, such as a limited number of observable neighbor nodes for each observer node, is anticipated in future work. 

\begin{acknowledgments}
This research was partially supported by JST, ERATO, Kawarabayashi Large Graph Project.
The empirical network data were downloaded from Stanford Large Network Dataset Collection~\cite{snap}.
T.H.'s work was partially supported by the Grant-in-Aid for Young Scientists (B) of JSPS (Grant No. 24740054).
Y.Y. was supported by JSPS Grant-in-Aid for Research Activity Start-up (Grant No.~24800082), and MEXT Grant-in-Aid for Scientific Research on Innovative Areas (Grant No.~24106003).
\end{acknowledgments}

\renewcommand{\theequation}{A.\arabic{equation}}
\section*{Appendix}
\subsection*{A. Analysis of $U$ for the clustered random graph}
To obtain analytical insights into the relationship between $C (\neq 0)$ and $U$,
we consider the clustered random graph (CRG) model~\cite{Newman2009,Miller2009}.
The CRG has two advantages: the $C$ value is controllable without additional structural manipulations (\eg, link rewiring)
and all the triangles are disjoint. 
Therefore, the structure of the CRG is determined by the joint degree distribution $p_{s,t}$ representing the probability that a node has $s$ dyadic links and $t$ triangle links~\cite{Newman2009}.
The degree distribution $p(k)$ can be derived as $p(k) = \sum_{s+2t = k} p_{s,t}$.

We calculate $U$ for random observer placement on a CRG using generation-function formalism~\cite{Newman2009}.
Let us define the generating function of $p_{s,t}$ by
$G_0(x,y) = \sum_{s,t} p_{s,t} x^s y^t.$
The degree distributions of a node connected to a dyadic link and triangle link are denoted by $q_{s,t} = (s+1)p_{s,t}/\langle s \rangle$
and $r_{s,t} = (t+1)p_{s,t} / \langle t \rangle$, respectively.
The generating functions of these distributions are given by
$G_q(x,y) = \sum_{s,t} q_{s,t} x^s y^t$ and $G_r(x,y) = \sum_{s,t} r_{s,t} x^s y^t$.
We refer to the nodes that are not observer nodes or their neighbors as unprotected nodes.
Let $u_q$ and $u_r$ be the probabilities that a node adjacent to a dyadic link and a triangle link, respectively, does not belong to the largest connected component
of unprotected nodes.
These probabilities obey the following recursive equations:
\begin{align}
u_q &= \phi + (1-\phi)G_q(u_q, u_r),\\
u_r &= \phi_\Delta + (1-\phi_\Delta)G_r(u_q, u_r),
\end{align}
where $\phi$ is the probability that a node is an observer node
and $\phi_\Delta \equiv 1 - (1-\phi)^2$ is the probability that at least one of two neighboring nodes connected by a triangle link is an observer.
Using $u_q$ and $u_r$, $U$ is obtained by
\begin{equation}
U = 1 - G_0(u_q, u_r).
\label{eq:U}
\end{equation}

\begin{figure}[t]
\centering
\includegraphics[width=0.5\hsize]{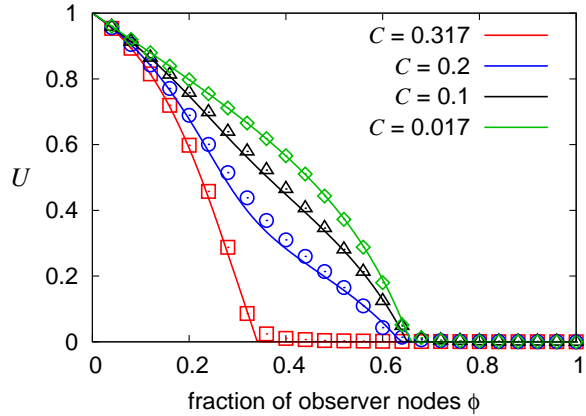}
\caption{
(Color online) The sizes of the largest connected component of unprotected nodes $U$ for random observer placement
are plotted as a function of $\phi$ on a clustered random graph with $N=10000$ and regular degree $k = 4$ for various values of $C$.
The symbols show $U$ obtained from the numerical simulations of random observer placement
(we take the average of $U$ for $100$ trials of random placement over ten network instances).
The lines show the theoretical solutions given by Eq.~\eqref{eq:U}. 
We set $C=0.317,$ $0.2,$ $0.1,$ and $0.017$ for the four lines from left to right.}
\label{fig:U_clusteredRRG_k4}
\end{figure}

In Fig.~\ref{fig:U_clusteredRRG_k4}, $U$ for a CRG with regular degree of four is plotted as a function of $\phi$ for various values of $C$.
In this case, $p_{s,t}$ consists of three elements, \ie, $p_{4,0}$, $p_{2,1}$, and $p_{0,2}$.
Theoretically, $C$ is given by $C = \left( p_{2,1} + 2p_{0,2}\right)/6$.
We realize different $C$ values by changing $p_{4,0}$ to $0,$ $0.35,$ $0.65,$ and $0.9$,
while keeping $p_{2,1} = 0.1$ and $p_{0,2} =  1 - p_{4,0} - p_{2,1}$.
Our numerical results (symbols) and theoretical solutions (lines) given by Eq.~\eqref{eq:U} agree well for all $C$ values.
In addition, if $\phi$ is fixed, $U$ decreases with $C$.
This result is consistent with our simulation results (Fig.~\ref{fig:R_RRG_lambda}), implying that random observer placement is more effective in highly clustered networks. 

\end{document}